\documentclass[twocolumn,showpacs,preprintnumbers,amsmath,amssymb]{revtex4}

\usepackage[dvipdfm]{graphicx}
\usepackage[dvipdfm]{color}
\usepackage{array, bm}
\usepackage{float}
\usepackage{layout}
\usepackage{amsmath, amssymb}
\usepackage{ulem}

\usepackage{ulem}
\usepackage{color}                        
\definecolor{SLvermilion}{cmyk}{0,.8,1,0} 
\definecolor{SLorange}{cmyk}{0,0.5,1.0,0}
\definecolor{SLblue}{cmyk}{1.0,0.5,0,0}

\begin{document}
\newlength{\pswlen}
\newlength{\pshlen}
\setlength{\pswlen}{7.0cm}
\setlength{\pshlen}{3.8cm}

\title{Weyl law for open systems with sharply divided mixed phase space}

\author{Akihiro Ishii}
\affiliation{Department of Physics, Tokyo Metropolitan University, Minami-Osawa, Hachioji, Tokyo 192-0397, Japan}
\author{Akira Akaishi}
\affiliation{Department of Physics, Tokyo Metropolitan University, Minami-Osawa, Hachioji, Tokyo 192-0397, Japan}
\author{Akira Shudo}
\affiliation{Department of Physics, Tokyo Metropolitan University, Minami-Osawa, Hachioji, Tokyo 192-0397, Japan}
\affiliation{Max-Planck-Institut f\"ur Physik Komplexer Systeme, N\"othnitzer Strasse 38, 01187 Dresden, Germany}
\author{Henning Schomerus}
 \affiliation{Department of Physics, Lancaster University, Lancaster LA1 4YB, United Kingdom}

\date{\today} 

\begin{abstract}
A generalization of the Weyl law to systems with a sharply divided mixed phase space is proposed. The ansatz is composed of the usual Weyl term which
counts the number of states in regular islands and a term associated with sticky regions in phase space. For a piecewise linear map, we numerically check
the validity of our hypothesis, and find good agreement not only for the case with a sharply divided phase space but also for the case where tiny island
chains surround the main regular island. For the latter case, a nontrivial power law exponent appears in the survival probability of classical escaping
orbits, which may provide a clue to develop the Weyl law for more generic mixed systems.

\end{abstract}


\maketitle

\section{Introduction}\label{sec:intro}

In closed quantum systems, a unit Planck cell supports a single eigenstate, and thereby the number of eigenstates up to a given energy $E$ is expected to
increase as ${\cal N}(E) \propto \mu(K(E)) / (2\pi \hbar)^d$, where $\mu(K(E))$ and $d$ denote the volume of the classical phase space $K(E)$
 and the dimensionality of
the system, respectively. 
A more precise mathematical estimation was made by Weyl on the asymptotic growth rate of ${\cal N}(E)$ for eigenmodes of the
Laplacian on bounded domains, which was originally provoked by the black body radiation problem \cite{Weyl12}, and analogous asymptotic laws in generic
quantum systems are now called the Weyl law.  The Weyl law is universal since the growth rate of eigenmodes is semiclassically determined only by the
volume of closed domains (for higher order length and curvature corrections see Ref.~\cite{BH78}), 
not by the nature of the underlying classical dynamics,
such as the integrability of the system.

It is natural to seek a generalization of the Weyl law to open systems in which Hamiltonians are no longer Hermitian and where spectra become complex
\cite{SZ}. Under the name of the fractal Weyl law it has been asserted that an analogous formula could be introduced for complex energies (resonances) of
open systems \cite{LSZ03,ST04}. Heuristic derivations together with mathematical arguments tell us that the Hausdorff dimension of the repeller in the
corresponding classical phase space
 characterizes the growth rate
of resonance energies \cite{LSZ03,ST04}. Numerical calculations \cite{LSZ03,ST04,Shep08}, and detailed mathematical analyses \cite{NZ05} then followed to
discuss the validity of the proposed formula.

In contrast to the original Weyl law in which all the invariant components in classical phase space, either regular or chaotic, equally contribute to the
density of states of eigenenergies, the corresponding Weyl law in open systems is rather subtle; it concerns the imaginary part of resonance states which
are necessarily linked to the finite-time dynamics, so may reflect the nature of underlying classical dynamics and its correspondence to quantum mechanics
as well \cite{ST04}. Therefore it is not straightforward to go beyond the situation that the classical system is strongly chaotic and the dynamics are 
homogeneous in phase space.

The aim of the present article is to argue for a possible extension of the Weyl law in more generic situations, especially in the case where the phase
space is composed of regular and chaotic components. In generic Hamiltonian systems, it is well known that regular and chaotic regions coexist in phase
space, and the motion along the border between them becomes sticky, which invokes long-time correlations in the dynamics \cite{Ka83,CShe84}. Although a
Weyl law for a mixed {\it closed} system was discussed recently \cite{BRLS11}, there is only limited understanding of whether or not an analog of the fractal
Weyl law exists in mixed open systems \cite{KS10,SGS10}.

Since geometrical structures in a mixed phase space become immensely complex in general, our strategy here is to take a simple system with sharply or
almost sharply divided phase space, for which the time scales associated with sticky motions are more controllable. We hypothesize a relation expected to
hold between the number of resonance states and the survival probability of sticky classical orbits, and then validate our ansatz in several typical
situations.

This work is organized as follows. In Sec.~\ref{sec:model} we describe a model of an open system with a sharply or almost sharply divided phase space, and
examine its classical escape dynamics.  In Sec.~\ref{sec:weyllaw} we formulate our ansatz for the Weyl law and validate its applicability to the resonances
of the quantized version of the model system. Section
\ref{sec:concl} contains our conclusions.

\section{Piecewise linear map and sticky motion in the open phase space}\label{sec:model}

Here we introduce a piecewise linear map with a mixed phase space and first investigate the classical escape. This will serve as the basis for the Weyl law
in the subsequent section.

\subsection{The case with a sharp border}

We consider the following piecewise linear map \cite{Wo81}:
\begin{eqnarray}\label{classical_map}
F: \left\{\begin{array}{l} p_{n+1} = p_n + kS(\theta_n) , \\
\theta_{n+1} = \theta_n + p_{n+1} \quad \mathrm{(mod ~1)}
,
\end{array}\right.
\end{eqnarray}
where $S(\theta)$ is a piecewise linear function given as
\begin{eqnarray}
S(\theta) = \left\{ \begin{array}{ll}
\theta & 0 \le \theta < 1/4, \\
-\theta + 1/2 &1/4 \le \theta < 3/4, \\
\theta - 1 &3/4 \le \theta < 1.\\
\end{array} \right.
\label{S_theta}
\end{eqnarray}
The dynamics is defined on a compact phase space $R$: $(\theta,p) \in [0,1)\times[-1,1]$, where periodic boundary conditions on $\theta$ and absorbing
boundary conditions on $p$ are imposed, respectively. 

\begin{figure}[htbp]
\begin{center}
\includegraphics[width=90mm]{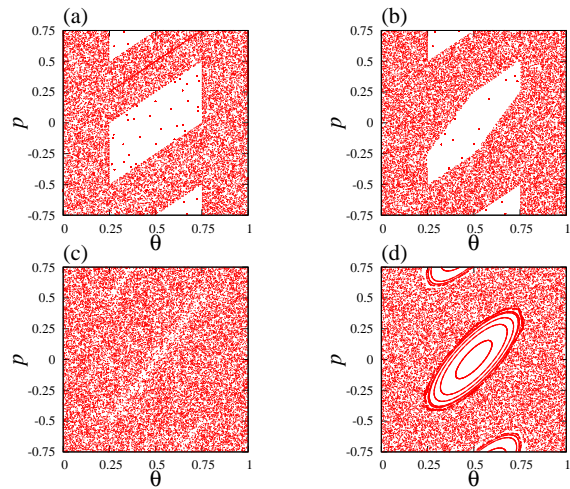}
\caption{(Color online)
Phase space of the map (\ref{classical_map}) for (a) $k=2.0$ [corresponding to $n=2$ in Eq. (\ref{k_condition})] and (b) $k=3.0$ $(n=3)$. For these values,
no hierarchical structures appear in phase space. Empty white regions are composed of periodic orbits. For (c) $k=4.0$, the regular region shrinks to a
family of stable periodic orbits. The case (d) $k=2.5$ does not satisfy the condition (\ref{k_condition}), and tiny hierarchical islands around the
regular region emerge. }
\label{Fig1}
\end{center}
\end{figure}

For the following specific parameter values $k$,
\begin{equation}
k = 2\Big(1 - \cos \frac{n - 1}{n}\pi \Big) \qquad (n = 1,2,...)
\label{k_condition}
\end{equation}
equipped with periodic boundary conditions on both coordinates $\theta$ and $p$, it was rigorously proved in \cite{Wo81} that the phase space is sharply
divided into a regular and a chaotic region. (Note, however, that the regular region is not a bundle of KAM curves as, e.g., in the mushroom billiards
\cite{Bu01}, but a set of periodic orbits.) Some phase space portraits are demonstrated in Fig. \ref{Fig1}. 

We first investigate the classical survival probability of the particles in the open phase space $R$. Figure \ref{Fig2} shows how the support of surviving
particles, initially distributed uniformly over the entire phase space, shrinks as time proceeds. Although the border between regular and chaotic regions
is sharp and no hierarchical structures appear, it was shown in \cite{AS09} that the orbits become sticky around the border, which causes a power-law
behavior in the survival probability: 
\begin{figure}[htbp]
\begin{center}
\includegraphics[width=80mm]{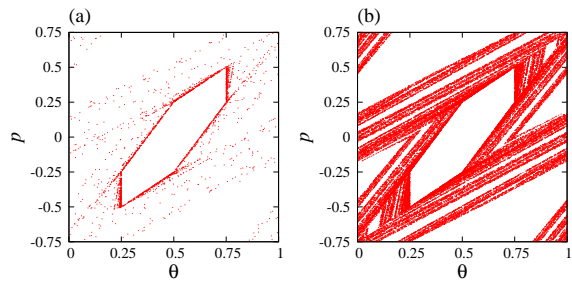}
\caption{(Color online)
The set of surviving points $(\theta_n,p_n)$ for $k=3.0$, after (a) $n=50$ iterations and (b) $n=10$ iterations, where $2\times10^5$ initial points are
distributed uniformly in the region $(\theta,p) \in [0,1) \times [-0.5,0.5]$ but outside the regular region. }
\label{Fig2}
\end{center}
\end{figure}
To see this, let $K^{\tau}$ be the points which remain in $R$ more than $\tau$ steps:
\begin{eqnarray}
K^{\tau} \equiv
\{ \,
(\theta,p) : F^{t}(\theta,p) \subset R ~~ {\rm for} ~t < \tau
\, \} .
\label{Ktau}
\end{eqnarray}
In Fig. \ref{Fig3}, we plot $\mu(K^{\tau})-\mu(K)$ as a function of $\tau$. Here $\mu(\cdot)$ denotes the area (Lebesgue measure) and $K \equiv
K^{\infty}$. 
\begin{figure}[htbp]
\begin{center}
\includegraphics[width=80mm]{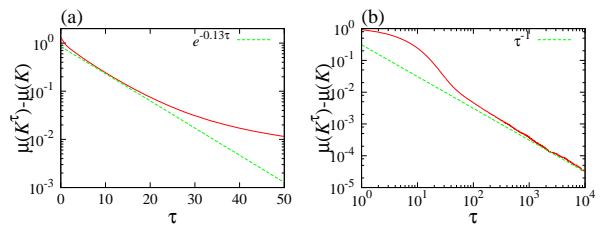}
\caption{(Color online)
Survival probability in the chaotic part in phase space, for the case $k=3.0$. (a) Normal-log and (b) log-log plot. $2^{20}$ points are initially
distributed uniformly over phase space, and the survival probability is obtained by counting the number of orbits remaining in $R$, up to $10^4$ steps.
Note that in panel (a), we observe exponential decay up to $\tau \lesssim 20$, while in panel (b) the decay follows a power law for $\tau \gtrsim 100$. }
\label{Fig3}
\end{center}
\end{figure}
The exponential decay in the short-time regime results from fast decay processes in the chaotic region, while the power-law
decay in the long-time regime is a typical characteristic of the sticky motion around the regular region \cite{AS09}. (The crossover takes place around
$\tau=\tau_c$, where $20\lesssim \tau_c\lesssim 100$.) Then we have
\begin{equation}
\mu(K^{\tau})-\mu(K) = C\tau^{-\nu} \qquad (\mathrm{for} \ \tau \gtrsim \tau_c).
\label{sticky_motion}
\end{equation}
Here the exponent and slope are obtained as $\nu = 1$ and $C \simeq 0.4$, respectively. Arguments to account for the exponential decay in the short-time
regime and a derivation of the power-law exponent $\nu = 1$ are found in \cite{AS09}.

\begin{figure}[htbp]
\begin{center}
\includegraphics[width=80mm]{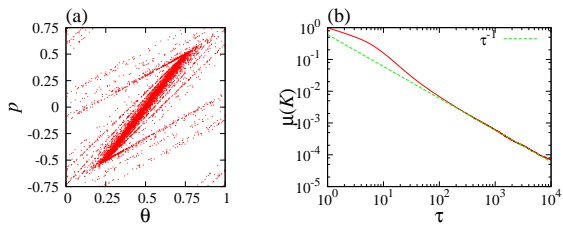}
\caption{(Color online)
(a) The set of surviving points $(\theta_n,p_n)$ for $k=4.0$ after $n = 50$ iterations, with $10^6$ initial points uniformly distributed
 over $R$.
(b) Survival probability for $k=4.0$. }
\label{Fig4}
\end{center}
\end{figure}

A similar behavior is observed in the case $k=4.0$. The corresponding survival probability is plotted in Fig.~\ref{Fig4}. In this case, since $\mu(K) =0$,
we have
\begin{equation}
\mu(K^{\tau}) = C\tau^{-\nu}\qquad (\mathrm{for} \ \tau \gtrsim \tau_c).
\end{equation}
The origin of the decay exponent $\nu = 1$ is the same as in the case $k=3.0$, and numerical fitting gives $C \simeq 1.0$. 

\subsection{The case with tiny islands}

\begin{figure}[htbp]
\begin{center}
\includegraphics[width=80mm]{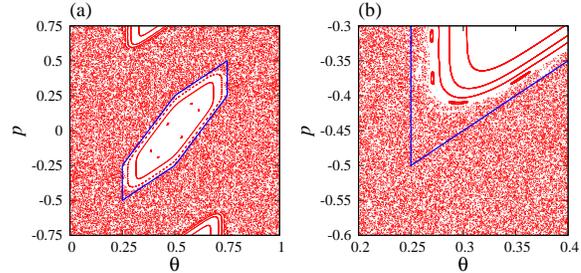}
\caption{(Color online)
(a) Phase space portrait  for $k=3.0$ and smoothed potential (\ref{smoothed_potential}) with $M=5$. The blue line represents the boundary between regular
and chaotic region in case of $M=\infty$. (b) Magnification around the boundary. Small islands appear due to the smoothing of the potential function. }
\label{Fig5}
\end{center}
\end{figure}

Next we examine the effect of small islands on the survival probability. Such islands are obtained when one replaces the piecewise linear function
$S(\theta)$ by
\begin{equation}\label{smoothed_potential}
S_{M}(\theta) = \frac{1}{2\pi} \sum_{l=0}^{M} \frac{(2l-1)!!}{(2l)!!(2l+1)} \sin^{2l+1} (2\pi \theta).
\end{equation}
The function $S_{M}(\theta)$ is obtained by expressing the piecewise linear function as $S(\theta)=1/2\pi \arcsin(\sin(2\pi\theta))$ and then truncating
the Taylor expansion of $\arcsin$ at the order $M$. The difference behaves as $|S(\theta)-S_M(\theta)| \sim \theta^{-(2M+1)}$, so $S_M(\theta)$ tends to
the original function $S(\theta)$ with increase of $M$. The smoothing induces tiny island chains surrounding the main regular island. As an example, we
present the case for $M=5$ in Fig.~\ref{Fig5}. 

\begin{figure}[htbp]
\begin{center}
\includegraphics[width=80mm]{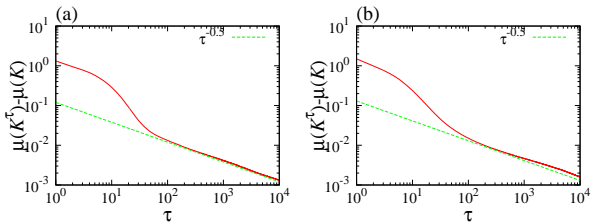}
\caption{(Color online)
Survival probability of the chaotic orbits for the smoothed potential (\ref{smoothed_potential}) with $M=100$ and (a) $k=3.0$, (b) $k=4.0$. }
\label{Fig6}
\end{center}
\end{figure}

For this moderately large value of $M$ islands become visible, which may in general give rise to a complicated behavior in the survival probability. For
$M=100$, on the other hand, the islands are tiny and invisible within the scale presented here, and the survival probability again decays algebraically, as
is shown in Fig.~\ref{Fig6}(a). However, the decay exponent $\nu$ changes from $\nu=1$ to $\nu=0.5$, even if the potential function is only slightly
smoothed. We also obtain $\nu=0.5$ for the case with $M=100$ and $k=4.0$ [see Fig.~\ref{Fig6}(b)].

The appearance of the observed exponent $\nu=0.5$ can be explained as follows: For large $M$, we may treat the sticky motion along the border between the
regular and chaotic regions in almost the same way as in the case with sharp boundaries. We can then apply the argument developed in
\cite{AMK06}, which is based on the relation between the survival probability
$\mu(K^\tau)$ and the distribution $P(\tau)$ of the time each orbit stays along the sticky layer. In each step, the orbit advances along the boundary by a
shift length $d$, which can be assumed to be proportional to a power of a distance $\varepsilon$ to the stable region, as $d\propto \varepsilon^r$, and the
time scale $\tau$ for which each orbit stays in the sticky region is then given as $\tau\sim \varepsilon^{-r}$. Denoting the distribution of initial points
in the sticky region by $p(\varepsilon)$, one can obtain the relation $P(\tau)\sim p(\varepsilon)|\frac{d\varepsilon}{d\tau}|$
\cite{AMK06}. Since we may assume the uniform distribution of initial points,
$p(\varepsilon)=\mathrm{const}$, due to ergodicity of the chaotic region, $P(\tau)\sim \tau^{-\frac{r+1}{r}}$ follows. The survival probability then turns
out to be $\mu(K^\tau)-\mu(K)\sim \tau^{-\frac{1}{r}}$.  In the case of a sharply divided phase space, $r=1$ \cite{AMK06}. One can also justify the
exponent for the smoothed case with tiny islands as $r=2$  in terms of perturbative calculations \cite{AS11}, which leads to $\nu=0.5$ unless fluctuations
due to the tiny islands strongly affect the motion in the sticky region. In the time regime that we here focus on, this condition is well fulfilled, and we
can thus predict how long it takes to exit from the phase space if $M$ is large enough.

\section{Weyl law for the piecewise linear map}\label{sec:weyllaw}

\subsection{Ansatz for the Weyl formula}

In order to arrive at the Weyl law for the open piecewise linear map with mixed phase space, we generalize considerations previously applied to fully
chaotic systems \cite{LZ02}. In analogy with the conventional Weyl law, we start with the relation
\begin{eqnarray}
W_{\gamma}(\hbar) \sim \frac{\mu(K^{\tau})}{(2\pi \hbar)^n} \qquad(\mathrm{as} \ \hbar \to 0)
\end{eqnarray}
Here $W_{\gamma}(\hbar)$ denotes the number of resonance states whose resonance width $\Gamma$ is less than $\gamma$, where the complex energy of each
resonance state is assumed to be expressed as $E - {\rm i}\Gamma$, and $\gamma \sim 1/\tau$. This heuristic formula is simply based on the hypotheses that
the number of resonance states with life time $1/\Gamma > 1/\gamma$ should be supported in the semiclassical limit by the set of classical orbits staying
in phase space for $T > \tau $ \cite{LZ02}. The crucial point of our analysis is to identify the correct cutoff scale for $\gamma$. In the chaotic case
this is determined by the inverse of the Ehrenfest time $\tau_E=(1/\lambda) \ln \hbar$, where $\lambda$ is the Lyapunov exponent \cite{ST04}. One then
obtains a fractal Weyl law with $W\propto \hbar^d$, where $d$ is related to the dimension of the fractal repeller (which is not an integer). This choice of
the cutoff cannot hold for a mixed phase space, where the chaotic component in phase space is not uniform.

In the present situation, as shown in the previous section, $\mu(K)$ obeys the power law (\ref{sticky_motion}) and $n=1$, so we have
\begin{equation}
\label{ansatz0}
W_{\gamma}(\hbar) \sim
\frac{1}{2\pi \hbar} \Bigl(
\mu(K) + C \gamma^{\nu} \Bigr)
\qquad(\mathrm{as} \ \hbar \to 0).
\end{equation}
In the semiclassical limit, the first term is supposed to be the number of regular states supported by the main island and the second one to the states
associated with the sticky zone. We here focus particularly on the sticky states, and introduce the notation
 $W_{\gamma}^{\mathrm{sticky}}(\hbar)$
as
\begin{equation}
W_{\gamma}^{\mathrm{sticky}}(\hbar) \equiv W_{\gamma}(\hbar) - \frac{\mu(K)}{2\pi \hbar}.
\label{wsticky_def}
\end{equation}

Here we put $\gamma = \alpha \hbar$, which leads to our ansatz
\begin{equation}
W_{\alpha \hbar}^{\mathrm{sticky}}(\hbar) \sim \frac{C\alpha^{\nu}}{2\pi}
\hbar^{\nu-1}.
\label{ansatz}
\end{equation}
This ansatz amounts to an Ehrenfest time $\tau_E \propto 1/\hbar$. Power-law Ehrenfest times for systems with a mixed phase space have been conjectured before
\cite{KS10}, but only in connection with situations with a hierarchical phase space (where islands of stability are surrounded by smaller islands). Here, we
examine the validity of this ansatz for the described system with a sharply divided phase space.

\subsection{Classification of resonances}

We now assess the validity of Eq. (\ref{ansatz}) by direct diagonalization of the quantum counterpart. The quantization procedure of the piecewise linear
map attaching absorbing boundaries and its smoothed version is presented in the Appendix. As in the case of the classical map, absorbing boundaries are set
at $p=-3/4$ and $3/4$, and $\hbar$ is given as $3/(2\pi N)$, where $N$ is the dimension of the Hilbert space.

\begin{figure}[htbp]
\includegraphics[width=85mm]{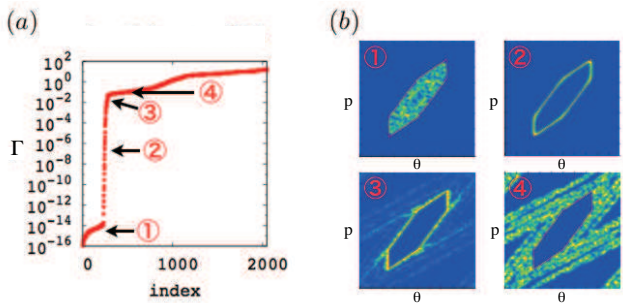}
\caption{(Color online)
(a) Quantum decay rates $\Gamma$ for $k=3.0$ and $N=4096$. The values of $\Gamma$ are arranged in the ascending order. (b) Four typical resonance states in
the Husimi representation. The numeric labels indicate the corresponding decay rate in panel (a). }
\label{Fig7}
\end{figure}

To start with, we identify several characteristic ranges of the quantum decay rate $\Gamma$, reflecting a variety of underlying classical dynamics. Figure
\ref{Fig7}(a) depicts the decay rate $\Gamma$ for $k=3.0$ arranged in ascending order. We may roughly classify four types of resonance states, numbered
as 1,2,3,4 in the figure:

\bigskip

\noindent
type 1 : localized in the main stable island.

\noindent
type 2 : localized at the edge of the main stable island.

\noindent
type 3 : localized in the sticky region.

\noindent
type 4 : extended in the chaotic region.

\bigskip

\noindent
If we count the number of states using the first term in (\ref{ansatz0}),  type 2 states turn out to follow the same Weyl law as the regular states (of
type 1).

The reason for the unusual appearance of type 1 regular states, which are not localized along invariant closed curves, but uniformly extended over the
whole regular region, is due to that, as mentioned, the orbits in the regular domain are not a bundle of KAM circles but the set of periodic points. Note
the similarity to the surviving classical orbits shown in Fig.~\ref{Fig2}, especially between the resonance of type 3 and the remaining points along the
border shown in Fig.~\ref{Fig2}(a), but also the resonance of type 4 and the orbits shown in Fig.~\ref{Fig2}(b).

\begin{figure}[htbp]
\includegraphics[width=85mm]{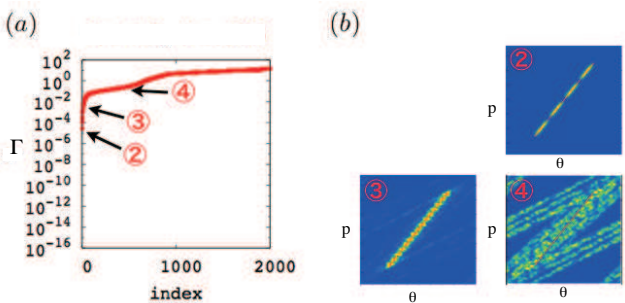}
\caption{(Color online)
(a) Quantum decay rates $\Gamma$ for $k=4.0$ and $N=4096$, arranged in the ascending order. (b) Three typical resonance states in the Husimi
representation. The labels again correspond to the indicated decay rates in panel (a). }
\label{Fig8}
\end{figure}

As shown in Fig.~\ref{Fig1}, the regular region for $k=4.0$ is just a set of fixed points, so types 1 and 2 states for $k=3.0$ are merged into type 2,
and sticky states are thus still numbered as type 3 in Fig.~\ref{Fig8}.

\subsection{Isolating resonances associated with sticky motion}

In order to see the validity of our ansatz (\ref{ansatz}), we first ensure that one can indeed separate the two terms in Eq. (\ref{ansatz0}). That is, we
isolate resonances associated with the sticky region from those associated with the regular region, and also check in which $\gamma$ range our ansatz
(\ref{ansatz0}) for the Weyl law associated with sticky motion holds. Here this is done for the case with a sharp border. In the next subsections, using the
results here obtained, we investigate the self-consistency of our argument, that is, the validity of our assumption $\gamma = \alpha \hbar$, and the range
of $\alpha$ where the ansatz (\ref{ansatz}) holds.

\begin{figure}[htbp]
\begin{center}
\includegraphics[width=85mm]{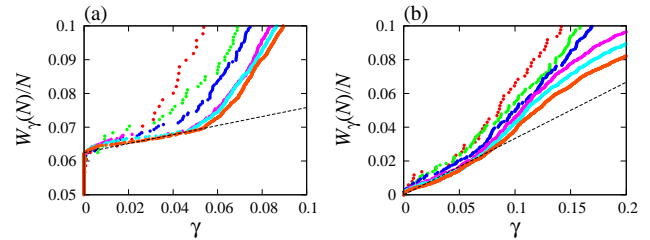}
\caption{(Color online)
$W_{\gamma}(\hbar)/N$ vs $\gamma$ for (a) $k=3.0$ and (b) $k=4.0$. The black dashed line expresses the formula (\ref{ansatz0}). The curves correspond to $N
= 512$ (red), 1024 (light green), 2048 (dark blue), 4096 (pink), 8192 (light blue), 16384 (orange). }
\label{Fig9}
\end{center}
\end{figure}

For the strictly piecewise linear map, we can rigorously evaluate the area of the regular island $\mu(K)$. For $k=3.0$, the regular island forms a polygon
with $\mu(K) = 3/16$, while $\mu(K) = 0$ for $k=4.0$. The coefficient $C$ appearing in (\ref{ansatz0}) has already been evaluated above, so inserting these
data into the formula (\ref{ansatz0}), we can check its validity by direct comparison to the numerical results for the quantum map. This is done in
Fig.~\ref{Fig9}, where we plot $W_{\gamma}(N)/N$ as a function of $\gamma$ for several values of $\hbar$, while the dashed line represents the right-hand
side of (\ref{ansatz0}).

For the range $\gamma \lesssim 0.05$, our ansatz agrees well with the numerical data, in both cases $k=3.0$ and $k=4.0$, and $W_{\gamma}(N)/N$ tends to
$\mu(K)/(2\pi\hbar)$ as $\gamma \to 0$. We also notice that with increasing $N$ the curves approach the expected line.

\subsection{The case with a sharp border }

The above results show that the separation of resonances associated with the sticky region from those associated with the regular region is a reasonable
assumption, and ranges of $\gamma$ actually exist where the ansatz (\ref{ansatz0}) holds. Now we check the consistency of the ansatz  (\ref{ansatz}) by
varying the free parameter $\alpha$ such that $\gamma$ is contained in the range of validity confirmed above.

\begin{figure}[htbp]
\begin{center}
\includegraphics[width=80mm]{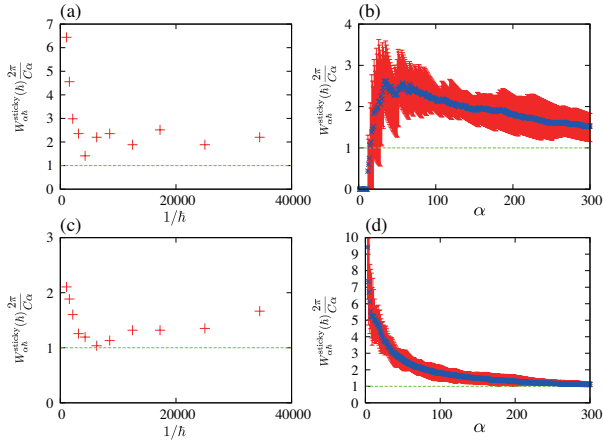}
\caption{(Color online)
Plot of $W_{\alpha \hbar}^{\mathrm{sticky}}(\hbar) \frac{2\pi}{C\alpha}$ as a function of $1/\hbar$ for (a) $k=3.0$ and (c) $k=4.0$. We put $\alpha = 100$
for $k=3.0$ and $\alpha=200$ for $k=4.0$. $W_{\alpha \hbar}^{\mathrm{sticky}}(\hbar) \frac{2\pi}{C\alpha}$ as a function of the parameter $\alpha$ for (b)
$k=3.0$ and (d) $k=4.0$, respectively. For each $\alpha$, we have calculated several values of $N$ (red curve) where $N$ is chosen such that $\gamma
\lesssim 0.05$ is satisfied. The blue line shows the average over $N$. If the ansatz (\ref{FWL_sticky_const}) holds, the function $W_{\alpha
\hbar}^{\mathrm{sticky}}(\hbar) \frac{2\pi}{C\alpha}$ should tend to the green line. }
\label{Fig10}
\end{center}
\end{figure}

We first investigate the strictly piecewise linear case, for which our ansatz can be written as
\begin{equation}
W_{\alpha \hbar}^{\mathrm{sticky}}(\hbar) \frac{2\pi}{C\alpha}
\sim 1 \qquad (\mathrm{as} \ \hbar \to 0).
\label{FWL_sticky_const}
\end{equation}
For $k=3.0$, we plot the left-hand side of Eq.~(\ref{FWL_sticky_const}) in Fig.~\ref{Fig10}(a), as a function of $1/\hbar$. The result demonstrates the
general validity of our ansatz. The deviations are small, and may be due to some localized states in the chaotic region which were counted as sticky ones.

Since $\alpha$ is a free parameter, we may vary $\alpha$ as long as the condition $\gamma \lesssim 0.05$ is satisfied. Figure \ref{Fig10}(b) is a plot of
$W_{\gamma}^{\mathrm{sticky}}(\hbar) \frac{2\pi}{C\alpha}$ for various $\alpha$. In this figure, $N$ is chosen such that the condition $\gamma = \alpha
\hbar \lesssim 0.05$ is satisfied. For example, within the range $0 < \alpha \le 300$, if  $1/\hbar \gtrsim 6000(N \gtrsim 3000)$, the condition $\gamma
\lesssim 0.05$ is fulfilled. However, we cannot take $\alpha$ too small since for a fixed $N$ the number of resonance states satisfying the condition
$\Gamma< \gamma$ decreases with decreasing $\alpha$, and finally becomes empty. Therefore, there is a lower bound of $\alpha$ for each $N$ such that a
sufficiently large number of resonance states with $\Gamma< \gamma$ exists. The deviation from the expected value (green line) observed in the small
$\alpha$ regime can therefore be attributed to the lack of resonance states for sufficient statistics. We  note that too large $\alpha$ also leads to
improper counting for $W_{\gamma}^{\mathrm{sticky}}$ since type 4 states (chaotic states) are included for a given $\hbar$. Except for such small and large
$\alpha$ regimes, we observe that our ansatz is a reasonably good one. Figure \ref{Fig10} also shows the validity of the ansatz for $k=4.0$.

\subsection{The case with tiny islands}

Next we examine the case with tiny islands along the border. In this case, the ansatz takes the form
\begin{equation}
W_{\alpha \hbar}^{\mathrm{sticky}}(\hbar) \sim \frac{C\alpha^{0.5}}{2\pi} \hbar^{-0.5}
 \qquad (\mathrm{as} \ \hbar \to 0).
\label{W_sticky_islands}
\end{equation}

\begin{figure}[htb]
\begin{center}
\includegraphics[width=55mm]{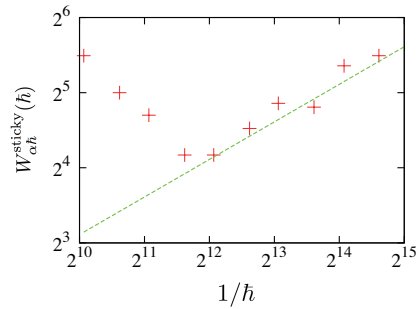}
\caption{(Color online)
Plot of $W_{\alpha \hbar}^{\mathrm{sticky}}(\hbar)$ as a function of $1/\hbar$ for $k=3.0$ and $\alpha = 100$. The green line represents $W_{\alpha
\hbar}^{\mathrm{sticky}} \simeq 0.27 \hbar^{-0.5}$. }
\label{Fig11}
\end{center}
\end{figure}

\begin{figure}[h]
\begin{center}
\includegraphics[width=85mm]{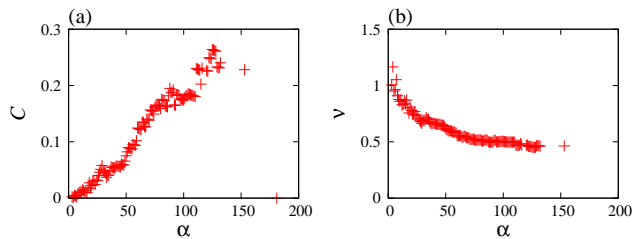}
\caption{(Color online)
For $k=3.0$, the assumed function $C(\alpha/6\pi)^{0.5} N^{0.5}$ is fitted to determine (a) $C$ and (b) $\nu$. }
\label{Fig12}
\end{center}
\end{figure}

For the $k=3.0$ case, we can only evaluate $\mu(K)$ numerically, so here we employ the numerical value $\mu(K^{10^6}) \simeq 0.178$, and $C \simeq 0.15$. As seen
in Fig. \ref{Fig11}, we again find that the numerical data agree well with the expectations once $\hbar$ is small enough. As shown in Fig.\ref{Fig12}, the
value of $C$ does not necessarily keep constant, but the exponent obtained by fitting the numerical data approaches the desired value $\nu \simeq 0.5$ [see
Fig.~\ref{Fig12}(b)].


\begin{figure}
\begin{center}
\includegraphics[width=55mm]{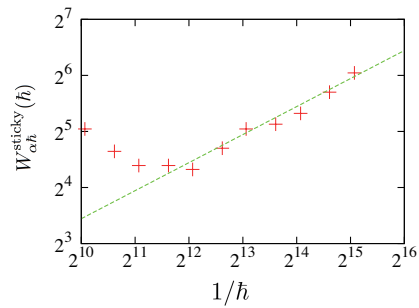}
\caption{(Color online)
Plot of $W_{\alpha \hbar}^{\mathrm{sticky}}(\hbar)$ as a function of $1/\hbar$ for $k=4.0$ and $\alpha = 100$. The green line represents
$W^{\mathrm{sticky}} \simeq 0.34\hbar^{-0.5}$. }
\label{Fig13}
\end{center}
\end{figure}

\begin{figure}[htb]
\includegraphics[width=85mm]{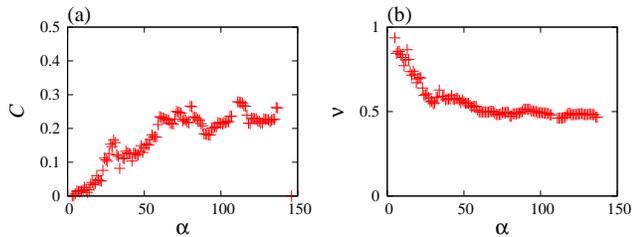}
\caption{(Color online)
For $k=4.0$, the assumed function $C(\alpha/6\pi)^{0.5} N^{0.5}$ is fitted to determine (a) $C$ and (b) $\nu$. }
\label{Fig14}
\end{figure}

In a similar way, we can also test the $k=4.0$ case. Using the numerical value $\mu(K^{10^6}) \simeq 0.0089$, we plot $W_{\alpha
\hbar}^{\mathrm{sticky}}(\hbar) \frac{2\pi}{C\alpha}$ as a function of $1/\hbar$ in Fig. \ref{Fig13}, and confirm that the ansatz works reasonably well for
the small $\hbar$ regime. This is further justified by fitting numerical data for various $\alpha$. We find that not only $\nu$ but also $C$ agree with
their classical values (Fig. \ref{Fig14}).


\section{Conclusions and Discussion}\label{sec:concl}

In this paper we have proposed a formula which generalizes the Weyl law to systems with a mixed phase space, and tested it numerically for a piecewise
linear map and its smoothed version. The formula is based on a simple classical-to-quantum correspondence for the states with finite lifetime, and is
composed of the classical measure of regular islands and the survival probability in the chaotic region. The former is associated with stable states
supported by regular classical islands, and the latter reflects classical sticky motion along the border between the regular and chaotic regions.

Classical dynamics in a mixed phase space is very complicated in general, and a single uniform time scale characterizing the dynamics does not exist. The
fractal Weyl law for chaotic systems, on the other hand, could be derived using the combination between classical-to-quantum correspondence, requiring
quasideterministic decay following classical decay process, and the classical survival time \cite{ST04}. Hence an analogous argument as for globally
hyperbolic systems cannot apply to generic mixed systems.

In addition, in the case where regular islands with positive measure coexist with the chaotic repellers in phase space, the Hausdorff dimension of the
whole invariant set is equal to 1. Thus, if we naively follow the original fractal Weyl law, the exponent characterizing the growth rate of the resonances
turns out to be unity, meaning that the fractal structure of chaotic repellers cannot reflect in the Weyl law. An idea of a $\lq\lq$fat fractal" has been
proposed to characterize such a situation
\cite{UF85}, and the Weyl law in the case where fat fractals appear in phase space
has been discussed in Refs. \cite{KS10,SGS10}.

Our strategy was to avoid the full complexity of a generic mixed phase space by taking a system with a sharply or almost sharply divided phase space, and
to see how the sticky dynamics affects the growth rate of resonance states. Even in such a simple setting, the time scale in which classical-to-quantum
correspondence holds is not obvious, and so there is no guarantee that one may identify the lifetime of quantum resonance states with classical survival
time.

Purely quantum effects such as localization or tunneling (diffraction in the sharp limit) must come into play when the effective Planck's constant is not
small enough. However, the observed good agreement of the power-law exponent in the proposed Weyl law not only in the sharply divided phase space but also
for the case with tiny islands provides significant resources for future investigations.

\section*{ACKNOWLEDGEMENT}
We thank A. Ishikawa, and S. Nonnenmacher and A. Tanaka for useful comments and discussions. We are also grateful to S. Tomsovic for helpful suggestions on classical-to-quantum
correspondence in mixed systems. This work is supported by Grant-in-Aid for Scientific Research (C) No. 21540934 from the Ministry of Education, Culture,
Sports, Science and Technology of Japan.

\appendix
\section*{Appendix: Quantization of the piecewise linear map with absorbing boundaries}

In this appendix we show how to quantize our piecewise linear map \eqref{classical_map} with absorbing boundaries and its smoothed version. The one-step
time evolution of a quantum state without absorbing boundaries is described by the unitary operator
\begin{eqnarray}
|\psi(n+1) \rangle = \hat U | \psi(n) \rangle ,
\end{eqnarray}
where
\begin{eqnarray}
\hat U =
\exp \Bigl[
-\frac{{\rm i}}{\hbar} \frac{\hat p^2}{2}
\Bigr]
\exp \Bigl[
-\frac{{\rm i}}{\hbar} V(\hat \theta)
\Bigr].
\end{eqnarray}
For the piecewise linear map (\ref{classical_map}), we set
\begin{eqnarray}
V(\theta) =
\left\{\begin{array}{ll}
-k\theta^2/2 & (0\le \theta < 1/4) \\ k(\theta^2/2 - \theta/2  + 1/16)& (1/4 \le \theta < 3/4)\\ -k(\theta^2/2 - \theta + 1/2) & (3/4 \le \theta < 1)
\end{array}\right.
\end{eqnarray}
and for its smoothed version,
\begin{eqnarray} \nonumber
V(\theta) &=& - \frac{k \cos{2\pi \theta}}{(2\pi)^2}\sum_{l=0}^{M} \frac{1}{(2l+1)^2} \\ &\times& \sum_{k=0}^{l}
\frac{(2l-2k-1)!!}{(2l-2k)!!} (\sin (2\pi \theta) )^{2l -2k}
,
\end{eqnarray}
which is obtained by integrating $S_M(\theta)$ given as (\ref{smoothed_potential}).

To introduce absorbing boundaries, we define the projection operator as
\begin{align}
\hat P(p) =
\begin{cases}
1 \quad (\text{$p$ $\in$ [-3/4,3/4]}),\\ 0 \quad (\text{$p$ $\notin$ [-3/4,3/4]}).
\end{cases}
\end{align}
Then we solve the eigenvalue equation
\begin{eqnarray}
\hat P(p) \hat U \Psi =
e^{-{\rm i}(E - {\rm i}\Gamma )} \Psi.
\end{eqnarray}
Here $E - {\rm i}\Gamma$ denotes the complex energy of the resonance states.

\end{document}